\documentclass[10pt,pre,aps,groupaddress]{revtex4}
\usepackage{epsfig,amsmath,graphicx,amssymb,overpic}
\usepackage{color}
\def\be{\begin{equation}}
\def\ee{\end{equation}}
\def\bee{\begin{eqnarray}}
\def\ene{\end{eqnarray}}
\def\bes{\begin{subequations}}
\def\ees{\end{subequations}}

\begin{document}

\title{Financial rogue waves}
\author{Zhenya Yan\footnote{Email address: zyyam@mmrc.iss.ac.cn}}
\affiliation{\small Key Laboratory of Mathematics Mechanization,
Institute of Systems Science, AMSS, Chinese Academy of Sciences,
Beijing 100190, China}

\vspace{2in}

\begin{abstract}
\baselineskip=14pt

{\bf Abstract.} \ We analytically present the financial rogue waves
in the nonlinear option pricing model due to Ivancevic, which is
nonlinear wave alternative of the Black-Scholes model. These rogue
wave solutions may be used to describe the possible physical
mechanisms for rogue
wave phenomenon in financial markets and related fields. \\


\end{abstract}

\vspace{2in}

\maketitle



\baselineskip=20pt

\section{Introduction}

Rogue waves have generated many marine misfortunes in the
oceans~\cite{RW1}. The New Year's wave or Draupner wave was
regarded as the first rogue wave recorded by scientific
measurement in North Sea. Recently, they were paid much attention
in order to understand better their physical mechanisms~\cite{RW1,
RW2, RW3, RW4, RW5, RW6, RW7, RW8}. Rogue waves are also known as
freak waves,  monster waves,  killer waves,  giant waves, or
extreme waves. The rogue wave phenomenon remain poorly understood.
It was not until 2007 that Solli {\it et al.}~\cite{ORW} first
observed the optical rogue waves in an optical fibre and found
that they could be used to stimulate supercontinuum
generation~\cite{exp0}. The basic solution (rogon) was first
presented by Peregrine~\cite{PS} to describe the rogue wave
phenomenon, which was known as by Peregrine soliton (or Peregrine
breather). Recently, the multi-rogon solutions were also presented
by using the deformed Darboux transformation in~\cite{ABC, ABC2}.
The matter rogue waves were realized by using the numerical
simulation~\cite{BRW} and the rogon-like solutions were also
found~\cite{ypla09}. In addition, the atmospheric rogue waves were
also presented~\cite{arw}.

To the best of our knowledge, there is no theoretical research for
the financial rogue waves (or financial crisis/storms) that have
been occurred (e.g. 1997 Asian financial crisis/storm) and are
taking place (e.g. the current global financial crisis/storm).

 Based on the the geometric Brownian motion (i.e. the
stochastic differential equation) $dS=\mu Sdt+\sigma dW(t)$
satisfied by the stock (asset) price $S$ and the It$\hat{\rm o}$
lemma~\cite{Ito}, the celebrated Black-Scholes linear partial
differential equation
 \bee
  \frac{\partial C}{\partial t}+\frac12\sigma^2 S^2\frac{\partial^2 C}{\partial S^2}
   +rS\frac{\partial C}{\partial S}-rC=0, \ene
was deduced~\cite{BS, BS2} , where $C\equiv C(S,t)$ is the values
of European call option on the asset price $S$ at time $t$, $\mu$
is the instantaneous mean return, $\sigma$  is the stock
volatility, $W$ is a Wiener process,  and $r$ is the risk-free
interest rate. In 1997, Merton and Scholes received the Nobel
Prize in Economy for their method to determine the price of a
European call option. But the model can not describe long-observed
features of the implied volatility surface.

 \section{Ivancevic option pricing model}

 Recently, Ivancevic, based on the modern adaptive markets
hypothesis due to Lo~\cite{Lo, Lo2} and Elliott wave market
theory~\cite{Elliot, Elliot2}, and quantum neural computation
approach~\cite{Ivan}, proposed a novel nonlinear option pricing
model (called the {\it Ivancevic option pricing model})
 \bee
  \label{nls}
   i\frac{\partial \psi(S,t)}{\partial t}=-\frac12\sigma\frac{\partial^2 \psi(S,t)}{\partial S^2}
    -\beta\left|\psi(S,t)\right|^2\psi(S,t), \ene
in order to satisfy efficient and behavioral markets, and their
essential nonlinear complexity, where $\psi=\psi(S,t)$ denotes the
{\it option-price wave function}, the dispersion frequency
coefficient $\sigma$ is the volatility (which can be either a
constant or stochastic process itself), the Landau coefficient
$\beta=\beta(r,w)$ represents the adaptive market potential. Some
periodic wave solutions of Eq. (\ref{nls}) have been
obtained~\cite{Ivan2}.

 \section{Financial rogue waves}

Here, based on the approach developed in ~\cite{ABC, ABC2}, we
show that the Ivancevic option pricing model~(\ref{nls}) also
possesses the financial multi-rogon (rogue wave) solutions, which
may be used to describe the possible formation mechanisms for
rogue wave phenomenon in financial markets. Here we give the first
two representative financial rogon solutions of the Ivancevic
option pricing model~(\ref{nls}).

 The financial one-rogon solution
of Eq.~(\ref{nls}) for the option-price wave function $\psi(S,t)$
by means of the complex rational functions of the stock price $S$
and time $t$ in the form
 \bee
 \label{solu1} \psi_1(S,t)=\alpha\sqrt{\frac{\sigma}{2\beta}}\left[1-\frac{4(1+i\sigma\alpha^2t)}
  {1+2\alpha^2(S-\sigma kt)^2+\sigma^2\alpha^4t^2}\right]
 \exp\big\{i[kS+\sigma/2(\alpha^2-k^2)t]\big\}, \ \ \sigma\beta>0 \
\  \ene
which involves four free parameters $\sigma,\, \beta, \, \alpha$
and $k$ to manage the different types of financial rogue wave
propagations whose intensity $|\psi_1(S,t)|^2$ is displayed in
Fig.~1 for the chosen volatility $\sigma=0.3$, adaptive market
potential $\beta=0.03$, the scaling $\alpha=2$ and the gauge $k=0,
-1.5$. Notice that time $t$ in Fig.~1 can be chosen to be negative
since the solution is invariant under the translation
transformation $t\rightarrow t+t_0$.

\begin{figure}
\begin{center}
{\scalebox{0.42}[0.5]{\includegraphics{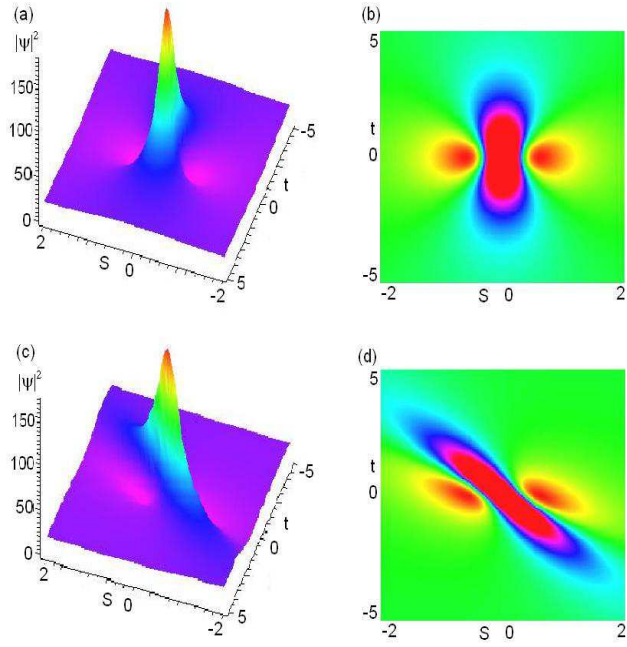}}}
\end{center}
\vspace{-0.2in} \caption{\small (color online). Rogue wave
propagations (left) and contour plots (right) for the intensity
$|\psi_1|^2$ of the one-rogon solution (\ref{solu1}) for
$\sigma=0.3,\, \beta=0.03,\, \alpha=2$. (a)-(b) $k=0$; (c)-(d)
$k=-1.5$. }
\end{figure}

\quad Moreover, the financial two-rogon solutions of
Eq.~(\ref{nls}) can be written as
 \bee \label{solu2}
\psi_2(S,t)=\alpha\sqrt{\frac{\sigma}{2\beta}}
 \left[1+\frac{P_2(S,t)+i\,Q_2(S,t)}{R_2(S,t)}\right]
 \exp\big\{i[kS+\sigma/2(\alpha^2-k^2)t]\big\}, \ \
\sigma\beta>0
  \ene
with these functions $P_2(x,t), \ Q_2(x,t)$ and $R_2(x,t)$ being
of  polynomial forms of  the stock price $S$ and time $t$
 \bee
 \nonumber
  \begin{array}{l}
 P_2(S,t)=\displaystyle \frac38-\frac{1}{2}\alpha^4(S-\sigma kt)^4-\frac32\sigma^2\alpha^6t^2(S-\sigma
 kt)^2 \vspace{0.08in}\cr
 \qquad\qquad  \displaystyle
 -\frac58\sigma^4\alpha^8t^4-\frac{3}{2}\alpha^2(S-\sigma kt)^2-\frac94\sigma^2\alpha^4t^2,
\vspace{0.1in}\cr
  Q_2(S,t)=\displaystyle -\frac12\sigma\alpha^2t\Big[\alpha^4(S-\sigma kt)^4+\sigma^2\alpha^6t^2(S-\sigma
  kt)^2  \vspace{0.08in}\cr
 \qquad\qquad  \displaystyle
 +\frac14\sigma^4\alpha^8t^4 -3\alpha^2(S-\sigma kt)^2+\frac12\sigma^2\alpha^4t^2-\frac{15}{4}\Big],
\vspace{0.1in}\cr
 R_2(S,t)=\displaystyle  \frac{3}{32}+\frac{1}{12}\alpha^6(S-\sigma kt)^6+\frac18\sigma^2\alpha^8t^2(S-\sigma kt)^4
 \vspace{0.08in}\cr
 \qquad\qquad  \displaystyle +\frac{1}{16}\sigma^4\alpha^{10}t^4(S-\sigma kt)^2+\frac{1}{96}\sigma^6\alpha^{12}t^6
\vspace{0.08in}\cr
 \qquad\qquad  \displaystyle
 +\frac18\alpha^4(S-\sigma kt)^4-\frac38\sigma^2\alpha^6t^2(S-\sigma kt)^2
 \vspace{0.08in}\cr
 \qquad\qquad  \displaystyle+\frac{9}{32}\sigma^4\alpha^8t^4
  +\frac{9}{16}\alpha^2(S-\sigma kt)^2+\frac{33}{32}\sigma^2\alpha^4t^2, \end{array} \ene
which contains four free parameters $\sigma,\, \beta, \, \alpha$
and $k$ to manage the different types of financial rogue wave
propagations whose intensity $|\psi_2(S,t)|^2$ is depicted in
Fig.~2 for the chosen volatility $\sigma=0.3$, adaptive market
potential $\beta=0.03$, the scaling $\alpha=0.8$ and the gauge
$k=0, -1.5$.
\begin{figure}[!h]
\begin{center}
{\scalebox{0.42}[0.5]{\includegraphics{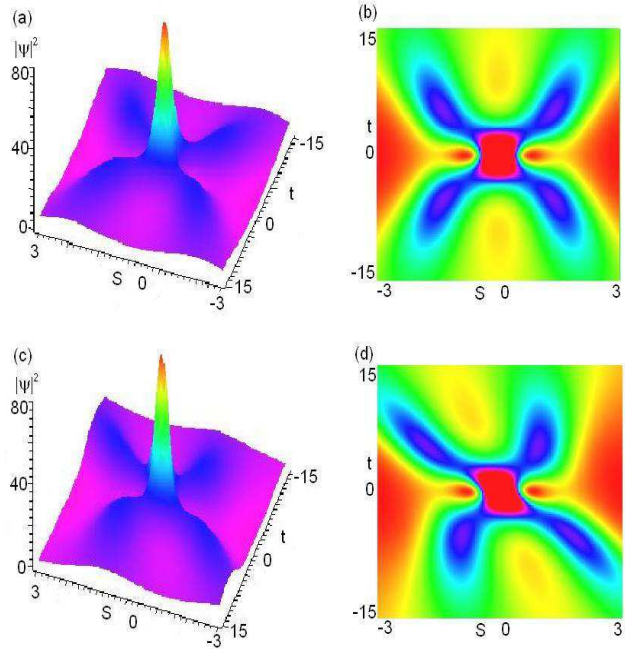}}}
\end{center}
\vspace{-0.2in} \caption{\small (color online).  Rogue wave
propagations (left) and contour plots (right) for the intensity
$|\psi_2|^2$ of the one-rogon solution (\ref{solu2}) for
$\sigma=0.3,\, \beta=0.03,\, \alpha=0.8$. (a)-(b) $k=0$; (c)-(d)
$k=-1.5$. }
\end{figure}

 \section{Conclusion}

 In conclusion, we have shown that the nonlinear option
pricing model (\ref{nls}) also possesses the analytical financial
one- and two-rogon solutions.  This may further excite the
possibility of relative researches and potential applications for
the financial rogue wave phenomenon in the financial markets and
related fields.

\vspace{0.02in}
 \noindent \textbf{\small Acknowledgement}
 The work
was supported by the NSFC60821002/F02.




\begin{thebibliography}{99}

{\small  \baselineskip=12pt

\bibitem{RW1} G. Lowton, New  Sci. 170 (2001) 28.

\bibitem{RW2} L. Draper, Mar. Obs. 35 (1965) 193.

\bibitem{RW3} C. Kharif, E. Pelinovsky, Eur. J. Mech. B (Fluids) 22 (2003) 603.

\bibitem{RW4} P. M\"uller, Ch. Garrett, A. Osborne, Oceanography 18 (2005) 66.

\bibitem{RW5} A. R. Osborne, Nonlinear Ocean Waves, Academic Press, New York, 2009.

\bibitem{RW6} C. Kharif, E. Pelinovsky, A. Slunyaev, Rogue
Waves in the Ocean, Observation, Theories and Modeling, Springer,
New York, 2009.

\bibitem{RW7} H. Tamura, T. Waseda, Y. Miyazawa,  Geophys. Res. Lett. 36 (2009) L01607.

\bibitem{RW8} K. Dysthe, H. E. Krogstad, P. M\"uller, Annu. Rev. Fluid Mech. 40 (2008) 287.



\bibitem{ORW} D. R. Solli, C. Ropers, P. Koonath,  B. Jalali, Nature 450 (2007) 1054.

 \bibitem{exp0} D. R. Solli, C. Ropers, B. Jalali, Phys. Rev. Lett. 101 (2008) 233902.


\bibitem{PS} D. H. Peregrine, J. Austral. Math. Soc. Ser. B 25 (1983) 16.

\bibitem{ABC} N. Akhmediev, A. Ankiewicz, J. M. Soto-Crespo, Phys. Rev. E 80 (2009) 026601.

\bibitem{ABC2}  N. Akhmediev, A. Ankiewicz, M. Taki, Phys. Lett. A 373 (2009) 675.

\bibitem{BRW} Yu. V. Bludov, V. V. Konotop, N. Akhmediev, Phys. Rev. A 80 (2009) 033610.

\bibitem{ypla09} Z. Y. Yan, Phys. Lett. A  374 (2010) 672.

\bibitem{arw} L. Stenflo and M. Marklund, arXiv:0911.1654.

\bibitem{Ito} K. It$\hat{\rm o}$,  Mem. Am. Math. Soc. 4 (1951) 1.

\bibitem{BS} F. Black and M. Scholes,  J. Pol. Econ. 81 (1973) 637.

\bibitem{BS2} R. C. Merton, J. Econ. Mana. Sci. 4 (1973) 141.


\bibitem{Lo} A. W. Lo, J. Portf. Manag. 30 (2004) 15.

\bibitem{Lo2} A. W. Lo,  J. Inves. Consult. 7 (2005) 21.


\bibitem{Elliot} A. J. Frost, R. R. Prechter,  Elliott Wave Principle:
Key to Market Behavior. Wiley, New York, (1978); (10th Edition)
Elliott Wave International, (2009)

\bibitem{Elliot2} P. Steven, Applying Elliott Wave Theory Profitably. Wiley,
New York, (2003)

\bibitem{Ivan} V. Ivancevic, T. Ivancevic, Quantum Neural Computation, Springer, New York,
2009.

\bibitem{Ivan2} V. Ivancevic, arXiv:0911.1834.


}


\end{thebibliography}
\end{document}